\documentclass[aps,floats,prl,
showpacs, twocolumn]{revtex4-1}

\usepackage{amsfonts,amsmath,mathrsfs} \usepackage{bm} \usepackage{bbm} \usepackage{dcolumn}
\usepackage{epsfig} \usepackage{latexsym}
\usepackage{multirow}

\begin{document}

\title{Transmission eigenvalues in random media with surface reflection}

\author{Xiaojun Cheng,$^{1,2}$ Chushun Tian,$^{3}$ and Azriel Z. Genack$^{1,2}$}

\affiliation{$^{1}$Department of Physics, Queens College,
The City University of New York, Flushing, NY 11367, USA\\
$^2$The Graduate Center, The City University of New York, New York, NY 10016 USA\\
$^3$Institute for Advanced Study, Tsinghua University,
Beijing 100084, China
}

\begin{abstract}
{\rm
The impact of surface reflection on the statistics of transmission eigenvalues
is a largely unexplored subject of fundamental and practical importance in statistical optics.
Here, we develop a first-principles theory and confirm numerically
that the distribution of transmission eigenvalues
of diffusive waves exhibits
a nonanalytic `transition' as the strength of surface reflection at one surface
passes through a critical value while that at the other is fixed.
Above the critical value, the highest transmission eigenvalue is strictly smaller than unity
and decreases with increasing internal reflection.
When the input and output surfaces are equally reflective,
the highest transmission eigenvalue is unity
and the transition disappears irrespective of the strength of surface reflection.
}
\end{abstract}

\date{\today}

\maketitle

The characteristics of transmission through random media are
determined by the statistics of the eigenvalues of the transmission matrix.
For example, universal conductance fluctuations of diffusive waves are due to
correlation between transmission eigenvalues \cite{Imry86},
while single parameter scaling of localized waves
reflects the dominance in the transmittance of
a single transmission eigenvalue with a log-normal distribution \cite{Anderson80}.
Powerful theoretical methods
have been developed to study the statistics of transmission eigenvalues
particularly in quasi-one-dimensional disordered waveguides \cite{Vasilev59,Dorokhov82,Dorokhov83,Dorokhov84,Mello88,Frahm95,Rejaei96,Zirnbauer04,Tian05}.
Only recently have measurements
of optical \cite{Popoff10} and microwave \cite{Genack12} transmission matrices been carried out and have
practical applications of the transmission matrix
in optical communications and imaging appeared to be feasible \cite{Popoff10,Mosk08,Mosk12}.

For diffusive waves in quasi-one-dimensional samples, the transmission
eigenvalues exhibit a bimodal distribution \cite{
Dorokhov84,Mello88,
Zirnbauer04,Tian05,Mello89,Nazarov94},
\begin{equation}\label{eq:15}
    \rho_0({\cal T})=
    \frac{\xi}{2L} \frac{1}{{\cal T}\sqrt{1-{\cal T}}}.
\end{equation}
and depends only upon the parameter $
\xi/L$ (see Ref.~\cite{Beenakker97} for a review of
the early status of this distribution). Here
$L$ is the sample length and $\xi$ is the localization length \cite{Efetov83a}.
The distribution of transmission eigenvalues is also bimodal
in higher dimensional diffusive samples \cite{Nazarov94}.
Although Eq.~(\ref{eq:15}) was originally predicted for electronic diffusive waves \cite{Dorokhov84},
it applies equally to classical waves such as light, sound, and microwave radiation
since it arises from coherent wave scattering.

The bimodal distribution was derived with reference to
electronic conductors coupled to ideal leads.
Additional aspects of random samples are
encountered in measurements of the transmission, such as
reflective interfaces (e.g., due to a mismatch in refractive index at the boundaries)
and absorption or amplification of radiation.
Understanding how these additional factors
affect the statistics of transmission eigenvalues
and the resultant transmission through random systems
will deepen our understandings of wave propagation in open complex media
and may guide efforts to
manipulate transmission for various applications by exciting selected eigenchannels.
The statistics of transmission is also affected by experimental considerations such as the inability
to measure the full transmitted field for all
incoming and outgoing channels \cite{Stone13} which has so far precluded the observation of the
the bimodal distribution \cite{Popoff10,Genack12}.

The average transmission
in the presence of internal reflection has been studied \cite{Lagendijk89,Zhu91,Genack93} using
the diffusion model or radiative transfer theory \cite{Morse53,Chandrasekhar,Rossum99}.
The impact of internal reflection on the average intensity
has been accounted for in a diffusion model. In this model,
the intensity inside the sample is found by solving the diffusion equation
in a region extending beyond the physical sample on each side by a length $z_b$
at which the intensity within the sample extrapolates to zero
\cite{Lagendijk89,Zhu91,Genack93,note_extrapolation}. In a sample with internal reflection
$R$, which may arise from the mismatch in refractive indices at the interface,
$z_b=0.7\ell (1+R)/(1-R)$ with mean free path $\ell$ \cite{Lagendijk89,Zhu91,Genack93}.
One might conjecture that the effect of surface reflection
upon the distribution of transmission eigenvalues could
be accounted for by replacing $L$ by the effective sample length $L+2z_b$ in Eq.~(\ref{eq:15}).
Whether the hypothesis is valid has not been explored.
It is clear however that the replacement: $L\rightarrow L+2z_b$ cannot be
generalized to localized waves.

Random matrix techniques \cite{Dorokhov82,Dorokhov83,Mello88}
are strictly valid in quasi one dimensions, but
not in high dimensions (e.g., the slab sample commonly used).
Though the Green's function approach used to derive the bimodal
distribution \cite{Nazarov94} is not restricted to low dimensions, it cannot be easily extended to localized waves.
Most importantly, it is
unclear how to include surface reflection in these techniques.
Recently, a powerful supersymmetric field theory
has been developed for classical wave propagation \cite{Tian08,Tian13}
and has provided an exact local
diffusion theory for localized waves in open random media \cite{Tian13,Tian10,Tian13a}.
The theory is valid for both diffusive and localized waves,
and applies to the slab as well as the quasi-one-dimensional geometries.
However, the challenging issue of the impact
of surface reflection on wave transport in open random media
has not been addressed by supersymmetric field theory so far.

\begin{figure}
 \begin{center}
 \includegraphics[width=8.0cm]{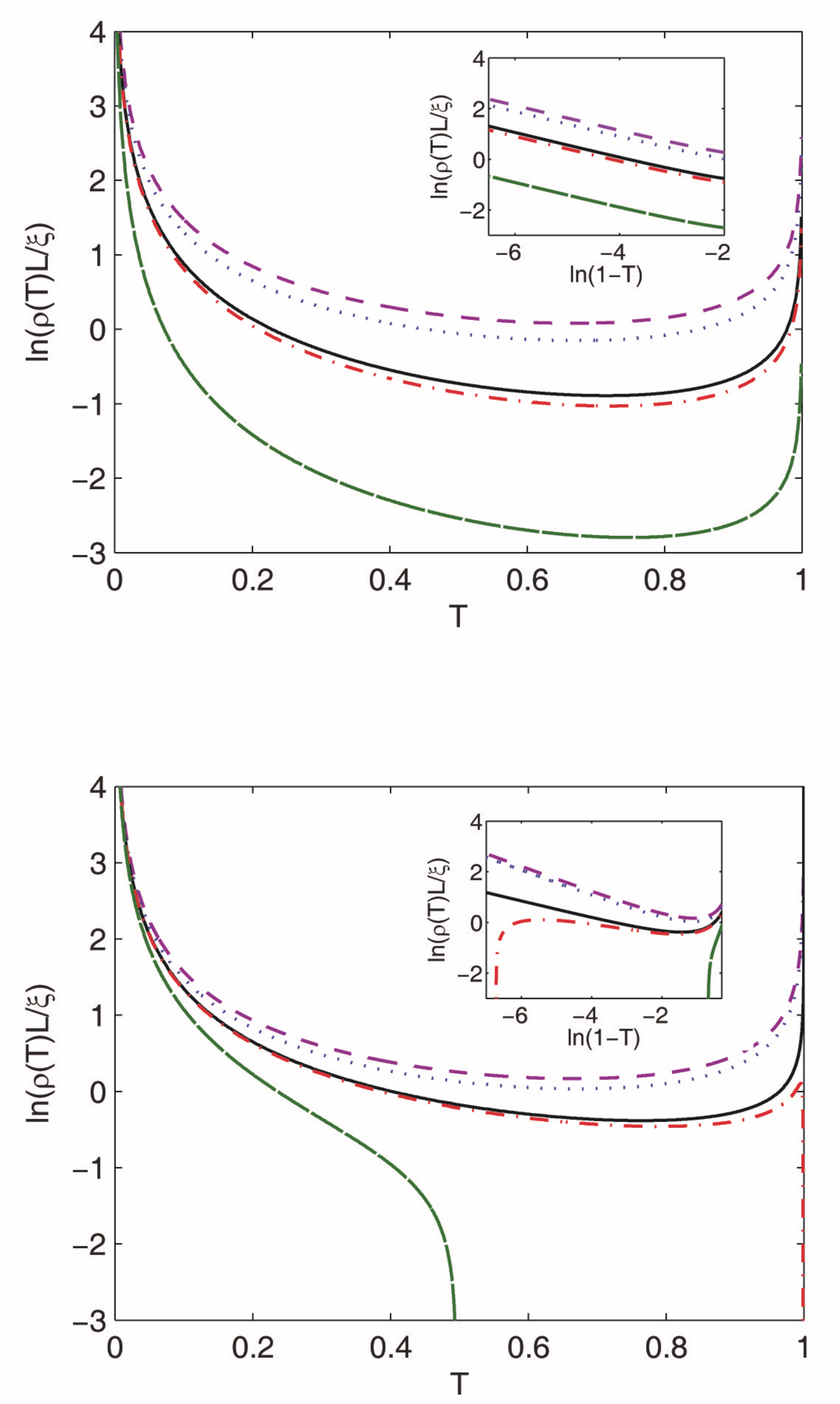}
\end{center}
 \caption{The analytical theory shows that
 as the internal reflection on the output surface increases,
 $\rho({\cal T})$ undergoes an `abrupt' change
 (`transition') at $\zeta=1$ (lower panel). This transition disappears
 when the internal reflection is the same on
 the input and output surfaces (upper panel). The values of $\zeta$ are
 $0.1,\, 0.25,\, 1,\, 1.2$, and $8$ from top to bottom. Inset: we
 re-plot the curves in the main panels in $\ln$-$\ln$ scale.}
 \label{fig:f_factor}
\end{figure}

In this Letter, we use
supersymmetric field theory to demonstrate that wave interference leads to
a nonanalytic `transition' of the distribution of transmission eigenvalues
as the internal reflection at one surface
passes through a critical value while
that at the other is fixed.
Above the critical value, the highest transmission eigenvalue is strictly smaller than unity
and decreases with increasing internal reflection.
When input and output surfaces are
equally reflective, the highest transmission eigenvalue is unity
and the transition disappears irrespective of the strength of surface reflection.
We find that although the distribution density
for high transmission eigenvalues is completely suppressed by strong internal
reflection at the input surface, surprisingly, it
is substantially enhanced by adding an identical reflector to the output.

We calculate the transmission eigenvalue distribution, $\rho({\cal T})$,
for diffusive samples where
$\ell\ll L \ll \xi$.
Because of numerous calculational subtleties, we first summarize
the main results. We express the deviation from Eq.~(\ref{eq:15})
via the factor
$f({\cal T})\equiv \frac{\rho({\cal T})}{\rho_{0}({\cal T})}$.
When the surface internal reflection $R$ is the same on
the input and output surfaces, $f$ is found implicitly from
(${\cal T}\equiv\cosh^{-2}\frac{\phi}{2}$)
\begin{eqnarray}
\phi=2{\rm arccosh} \frac{\pi\zeta f}{\cos\frac{\pi f}{2}}
+\frac{\sin\frac{\pi f}{2}}{\zeta}\sqrt{\left(\frac{\pi\zeta f}{\cos\frac{\pi f}{2}}\right)^2-1},
\label{eq:12}
\end{eqnarray}
where the parameter $\zeta\equiv z_b/L$. Equation (\ref{eq:12})
yields a family of curves of $\rho({\cal T})$ for different values of $\zeta$,
which are shown in the upper panel
of Fig.~\ref{fig:f_factor}. When only the output surface is reflective, $f$ is found from
\begin{eqnarray}
\phi={\rm arccosh} \frac{\pi\zeta f}{\sin \pi f}
-\frac{\cos \pi f}{\zeta}\sqrt{\left(\frac{\pi\zeta f}{\sin \pi f}\right)^2-1}.
\label{eq:16}
\end{eqnarray}
The plots for different values of internal
reflection are shown in the lower panel of Fig.~\ref{fig:f_factor}.
Equations (\ref{eq:12}) and (\ref{eq:16}) show that $f({\cal T})$
is governed by a single parameter $\zeta$.

For weak surface interactions, $z_b \ll L$,
the two cases are qualitatively similar:
Equations (\ref{eq:12}) and
(\ref{eq:16}) both reduce to $f({\cal T})=L/L_{\rm eff}$.
(Note that such uniform suppression breaks down for small ${\cal T}$, as
required by the normalization.)
Here the effective sample length $L_{\rm eff}$ is
$(L+2z_b)$ when the input and output surfaces are equally reflective and
$(L+0.7\ell+z_b)$ when only the output surface is reflective.
In the latter case, $0.7\ell$ arises from the transparent
input surface and $z_b$ from the reflective
output surface \cite{Morse53}. As expected, the distribution $\rho({\cal T})$
for relatively transparent samples differs from Eq.~(\ref{eq:15}) only by the overall factor,
i.e., $\xi/L\rightarrow \xi/L_{\rm eff}$.

\begin{center}
\begin{table}[htbp]
\newcommand{\tabincell}[2]{\begin{tabular}{@{}#1@{}}#2\end{tabular}}
\centering
\caption{\label{comparison} Asymptotic behavior of $\rho({\cal T})$ at ${\cal T}\rightarrow 1$.}
\begin{tabular}{c|c|c}
\hline\hline
\multirow{2}{*}{\tabincell{c}{location of \\ internal reflection}}
& \multirow{2}{*}{\tabincell{c}{input and output \\ surface}} & \multirow{2}{*}{\tabincell{c}{input or output \\ surface}} \\
&&\\
\hline
$\zeta<1$&\multirow{3}{*}{\tabincell{c}{$\sim(1-{\cal T})^{-\frac{1}{2}}$}}
&$\sim(1-{\cal T})^{-\frac{1}{2}}$\\
\cline{1-1}
\cline{3-3}
$\zeta=1$&&$\sim(1-{\cal T})^{-\frac{1}{3}}$\\
\cline{1-1}
\cline{3-3}
$\zeta>1$&&$0$\\
\hline
\end{tabular}
\label{tab:1}
\end{table}
\end{center}

For strong surface interactions,
$z_b\gtrsim L$,
the deviations from the bimodal distribution are very different for the two cases,
as shown in Fig.~\ref{fig:f_factor} and as summarized in Table~\ref{tab:1}.
In the case of internal reflection only at a single interface,
the asymptotic behavior of
$\rho({\cal T})$ for ${\cal T}\rightarrow 1$ undergoes
an abrupt change (`transition') as $\zeta$ passes through unity:
the singularity of $\rho({\cal T}\rightarrow 1)$ changes from $(1-{\cal T})^{-\frac{1}{2}}$ for $\zeta>1$
to $(1-{\cal T})^{-\frac{1}{3}}$ for $\zeta=1$,
while $\rho({\cal T}>{\cal T}_{\rm max})=0$ for
$\zeta>1$. The highest transmission eigenvalue ${\cal T}_{\rm max}$ decreases with $\zeta$
and ${\cal T}_{\rm max}\sim \zeta^{-1}$
for $\zeta\gg 1$. The same results are found
when internal reflection is present only at the input
instead of the output interface. The transition,
however, does not occur when both surfaces are equally reflective. In this case,
as internal reflection increases, the
asymptotic behavior does not change, i.e.,
$\rho({\cal T}\rightarrow 1)\sim (1-{\cal T})^{-\frac{1}{2}}$, aside from an overall factor.

\begin{figure}
 \begin{center}
 \includegraphics[width=8.0cm]{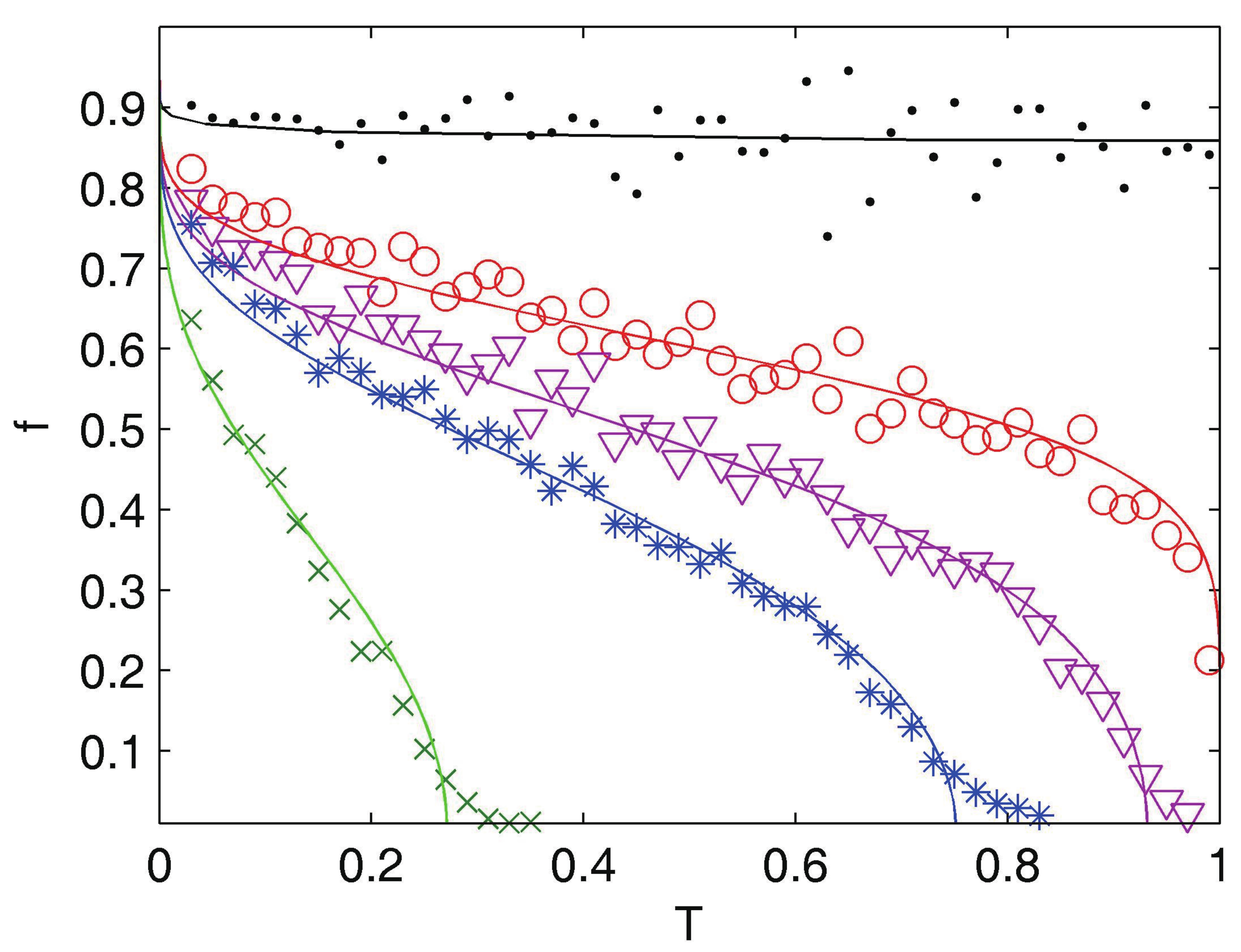}
\end{center}
 \caption{The numerical results and the analytic prediction of Eq.~(\ref{eq:16})
 are in good agreement. This confirms the transition in the transmission eigenvalue statistics.
 The fit with the single parameter $\zeta$ gives $\zeta=0.16,\, 0.93,\, 2.2,\, 4.0$,
 and $17.2$ from top to bottom. }
 \label{fig:numerical}
\end{figure}

The behavior of $f({\cal T})$ predicted by Eq.~(\ref{eq:16})
is confirmed in numerical experiments on
wave transport through disordered waveguides
using the recursive Green's function method. A scalar wave
is launched into a diffusive sample with $\xi/L\approx 16$, which is an
$1800 \times 600$ rectangular lattice.
The lattice spacing is unity (in units of the inverse wavenumber) with the wave velocity in the surrounding air set to unity.
The squared refractive index at
each site fluctuates independently around
the air background value, taking random values over the interval $[0.7,1.3]$.
To create internal reflection, we add an additional
layer of thickness $2$ lattice spacings and constant refractive index at the output of the sample.
Simulations were carried out for refractive indices of the surface layer of
$1.8, 2, 2.1, 2.2$, and $2.5$.
For each of these values, we found the $200$ transmission eigenvalues in
$3,000$ disordered configurations.
The results, shown in Fig.~\ref{fig:numerical}, are in good agreement
with the analytical results of Eq.~(\ref{eq:16}).
$\zeta$ is treated as the single fitting parameter
and is found to be $ 0.16,\, 0.93,\, 2.2,\, 4.0$,
and $17.2$ from top to bottom. Note that the appearance of data
points giving nonzero density above ${\cal T}_{\rm max}$ is
due to finite channel number in numerical experiments.

If the internal reflection at the input surface is fixed at some nonzero value,
the transition still occurs as the internal reflection at
the output surface passes through a certain critical value whose detailed form
[see Supplementary Materials (SM)] depends on the internal reflection at
the input surface. Therefore, it is a large imbalance in the internal reflection at the input
and output surfaces that leads to ${\cal T}_{\rm max}<1$.
Importantly, although
$\rho({\cal T}\geq {\cal T}_{\rm max})=0$ for
strong internal reflection on the input surface
(the $\zeta=8$ curve in Fig.~\ref{fig:f_factor}, lower panel), surprisingly,
$\rho({\cal T})$ for high values of ${\cal T}$ is substantially enhanced
when we add an identical reflector to the output
(the $\zeta=8$ curve in Fig.~\ref{fig:f_factor}, upper panel).
This implies that for ${\cal T}_{\rm max}<1$
wave functions are sensitive to the boundary
conditions and thereby must be delocalized.
Thus, this result is a coherent effect for diffusive waves and
is related to neither weak nor strong localization.
Rather, it resembles a well-known quantum mechanical phenomenon. That is,
although a single barrier prohibits wave transmission,
when an identical barrier is added, the system can be perfectly transmitting.
Eq.~(\ref{eq:16}) coincides with a result
for a completely different context \cite{Nazarov94,Beenakker94}, in which
a normal metal is coupled to another normal metal or to a superconductor
via a {\it single} barrier. However, the transition here has
nothing to do with the localization-delocalization transition of the wave function
suggested in Ref.~\cite{Nazarov94}, as shown above. Indeed,
we use the analytical result of $\rho({\cal T})$ (Fig.~\ref{fig:f_factor}, lower panel)
to numerically compute $\int_0^1 d{\cal T} {\cal T}\rho({\cal T})$
namely the average conductance (see SM for details). We
find that for $\zeta$ ranging from $10^{-1}$ to $10^3$, the average conductance
is $\xi/L_{\rm eff}$, i.e., Ohm's law is valid no matter whether
internal reflection is weak or strong.


Having summarized the main results, let us
present an outline of the microscopic theory. Because the results above are
valid irrespective of time-reversal symmetry,
for simplicity we focus on unitary systems for which this symmetry is broken.
We start from the case of identical values of
internal reflection on the
input and output surfaces. The transmission eigenvalue density
may be obtained from the function
$F(\phi)\equiv -\frac{i}{2}\sinh\phi \langle
    {\rm tr}[{\bf t}{\bf t}^\dagger/
    (1+\sinh^2(\phi/2){\bf t}{\bf t}^\dagger)]\rangle$
since
$\rho({\cal T}) =\frac{1}{2\pi}[F(\phi+i\pi)+F^*(\phi+i\pi)]
\frac{d\phi}{d{\cal T}}$,
where $\phi$ is understood as $\phi-i\delta$
with $\delta$ a positive infinitesimal,
and $\langle\cdots\rangle$ represents the average over Gaussian disorder.
The function $F(\phi)$ can be cast into a functional integral over the supersymmetric
field $Q(x)$ (see Ref.~\cite{Tian13} for a review of the applications
of supersymmetric field theory \cite{Efetov97}
to classical waves in open random media). Details of the calculation
are presented in SM.
We find
\begin{eqnarray}
F(\phi)&=& -\frac{i\xi}{2}
\int_{(2z_b Q\partial_x Q+[Q,\Lambda])|_{x=0}=0}^{(2z_b Q\partial_x Q - [Q,\Gamma])|_{x=L}=0} D[Q]\nonumber\\
&& \times (Q\partial_x Q)^{21}_{\rm bb}|_{
x=0,\theta=i\phi} e^{-\frac{\xi}{8}
\int_0^L dx
{\rm str} (\partial_x Q)^2},
\label{eq:7}
\end{eqnarray}
where `str' is the supertrace. $Q
$ is a $4\times 4$ supermatrix
defined on both the advanced-retarded (`ar')
and the fermionic-bosonic (`fb') sector.
More precisely, $Q=T^{-1}\Lambda T$ and the supermatrix $T$
takes the supermatrix value from the coset space
$U(1,1|2)/U(1|1)\otimes U(1|1)$. The crucial difference between Eq.~(\ref{eq:7}) and
the expression for disordered electronic wires coupled to ideal leads given in Refs.~\cite{Zirnbauer04,Rejaei96,Tian05}
is the radiative boundary condition at the interfaces, $x=0,L$ \cite{Tian13,Tian08}.
In Eq.~(\ref{eq:7}), $\Lambda$ and $\Gamma$ are constant supermatrices.
The former is ${\rm diag} (\mathbbm{1}^{\rm fb},\,-{\mathbbm{1}}^{\rm fb})^{\rm ar}$ and
the latter is
$$
{\rm diag}\!\left(\!
                   \left(
\begin{array}{cc}
                        \cos\theta & -i\sin\theta \\
                        i\sin\theta & -\cos\theta \\
                      \end{array}
                    \right)^{\rm ar},
\left(
                      \begin{array}{cc}
                        \cosh\phi & \sinh\phi \\
                        -\sinh\phi & -\cosh\phi \\
                      \end{array}
                    \right)^{\rm ar}\right)^{\rm fb},
$$
where $0<\theta<\pi$ and $\phi>0$.

The functional integral in Eq.~(\ref{eq:7}) is dominated
by fluctuations around field configurations
which solve the saddle point equation
$\partial_x (Q\partial_xQ)=0$ implemented with the radiative boundary conditions, i.e.,
$(2z_b Q\partial_x Q+[Q,\Lambda])|_{x=0}=0$ and $(2z_b Q\partial_x Q - [Q,\Gamma])|_{x=L}=0$.
The solution of $Q(x)$ has the same structure as $\Gamma$
except for the replacements: $\phi \rightarrow \Phi(x),
\theta \rightarrow \Theta (x)$\,. Substituting these into
the saddle point equation and the boundary conditions gives
\begin{eqnarray}
\partial_x^2 \Phi = \partial_x^2 \Theta = 0,
\label{eq:10}
\end{eqnarray}
satisfying the boundary conditions, i.e.,
\begin{subequations}
\begin{equation}\label{eq:46}
    z_b\partial_x \Phi -\sinh \Phi
    =z_b \partial_x \Theta -\sin \Theta=0,
\end{equation}
for $x=0$ and
\begin{equation}\label{eq:47}
    z_b\partial_x \Phi +\sinh (\Phi-\phi)
=z_b\partial_x \Theta +\sin (\Theta-\theta)=0,
\end{equation}
\end{subequations}
for $x=L$. The solution has the form $\Phi(x)=C_\phi x/L + \phi_0,\,
\Theta (x)=C_\theta x/L + \theta_0$, where $C_{\phi,\theta}\in \mathbbm{R}$,
$\phi_0>0$ and $0<\theta_0<\pi$. It is well defined for arbitrary
$z_b$ and converges to the limiting case of $z_b=0$ \cite{Zirnbauer04}. As a result
(see SM
for derivations),
\begin{eqnarray}
  \zeta C_\phi=-\sinh \frac{C_\phi-\phi}{2},\quad
  \zeta C_\theta=-\sin \frac{C_\theta-\theta}{2}
\label{eq:14}
\end{eqnarray}
once $\phi_0$ and $\theta_0$ are eliminated from Eqs.~(\ref{eq:46}) and (\ref{eq:47}).

Upon substituting $\Phi(x),\Theta(x)$ into the action
$\frac{\xi}{8}\int_0^L dx {\rm str}(\partial_x Q)^2$, we find
the saddle point action $\sim \frac{\xi}{L}(C_{\theta=i\phi}^2+C_\phi^2)$.
A set of saddle point actions results since the second equation
in (\ref{eq:14}) has a family of solutions. Taking
the analytic continuation of the second equation in (\ref{eq:14}), we find $C_{\theta=i\phi}=iC_\phi$.
As a result, the smallest saddle point action is zero, and is
gapped from the others by an action
$\sim {\cal O}(\frac{\xi}{L})$. Since we are interested in diffusive samples, $\xi/L\gg 1$, we
need to consider only the saddle point configuration
leading to a vanishing action and fluctuations around this.
Upon integrating out
fluctuations, a functional superdeterminant which is
unity at $\theta=i\phi$ results, thanks to the supersymmetry. We thereby obtain
$F(\phi)=-\frac{i}{2}\frac{\xi}{L}C_\phi$. [In deriving this result we
ignore fluctuations in the pre-exponential factor of Eq.~(\ref{eq:7})
since they only give rise to corrections of lower order.]

Supposing $z_b=0$, (This limit, however, cannot be obtained in
reality since $z_b \geq 0.7 \ell$ \cite{Morse53}.)
Eq.~(\ref{eq:10}) can be easily solved, giving
$C_\phi=\phi$. As a result,
$F(\phi)=-\frac{i}{2} \frac{\xi}{L}\phi$. From this, we obtain
$\rho({\cal T})=\frac{\xi}{2L}\frac{d\phi}{d{\cal T}}$, namely the bimodal
distribution (\ref{eq:15}). Thus, it is clear that this distribution
excludes interface effects. For weak internal reflection, $\zeta
\ll 1$, we may
expand the $\sinh$ and sine functions in Eqs.~(\ref{eq:14})
near zero. Keeping the expansion up to first order, we
obtain $f
\equiv (C_{\phi+i\pi}-C_{\phi-i\pi})/(2i\pi)=L/L_{\rm eff}$ with $L_{\rm eff}=L+2z_b$.
This gives the distribution
which is essentially the same as (\ref{eq:15}) except for the replacement
$L \rightarrow L_{\rm eff}$.
This same substitution gives the average transmission in the case of weak internal reflection
\cite{Lagendijk89,Zhu91,Genack93,Rossum99}. We see that the
effective role of weak internal reflection
indeed is to extend the sample length by $2z_b$.

Using the first equation
in (\ref{eq:14}), we can prove Eq.~(\ref{eq:12}) (see SM
for details).
Setting $\phi=0$ in Eq.~(\ref{eq:12}), we obtain $\pi\zeta f(1)
=\cos\frac{\pi f(1)}{2}$. This equation for $f(1)$ has a unique solution
in the
interval $(0,1)$ for arbitrary $\zeta$. Thus, $\rho({\cal T}\rightarrow 1)=
\frac{\xi}{2L}\frac{f(1)}{\sqrt{1-{\cal T}}}$, i.e., the amplitude of the peak of
the distribution (\ref{eq:15}) as ${\cal T}\rightarrow 1$ is suppressed, but the
square-root singularity does not change.

We now repeat the procedure above for the case in which
only the output surface ($x=L$) is reflective.
The key difference is that Eq.~(\ref{eq:14}) is replaced by
\begin{subequations}
\begin{equation}
\zeta C_\phi=-\sinh \left[\left(1+0.7\ell/L\right)C_\phi-\phi\right],
\label{eq:42}
\end{equation}
\begin{equation}
\zeta C_\theta=-\sin \left[\left(1+0.7\ell/L\right)C_\theta-\theta\right].
\label{eq:43}
\end{equation}
\end{subequations}
For $\zeta\ll 1$, we find $\rho({\cal T})=
    \frac{\xi}{2L_{\rm eff}} \frac{1}{{\cal T}\sqrt{1-{\cal T}}}$
with $L_{\rm eff}=L+0.7\ell +z_b$.
Again we see that the impact of weak internal reflection ($\zeta\ll 1$)
is to extend the sample length: $0.7\ell$ is the extension on the transparent
input surface and $z_b$ on the reflective
output surface. In general, using Eq.~(\ref{eq:42}), we are able to prove Eq.~(\ref{eq:16})
(see SM
for details).
For $\zeta<1$, we have
$\rho({\cal T}\rightarrow 1)=
\frac{\xi}{2L}\frac{f(1)}{\sqrt{1-{\cal T}}}$
as we found above, and the coefficient $f(1)$
now satisfies $\pi\zeta f(1)
=\sin \pi f(1)$. For $\zeta=1$, we
obtain from Eq.~(\ref{eq:16}) $\phi \sim f^3$ for $\phi\rightarrow 0$,
giving
distinct singularity,
$\rho({\cal T}\rightarrow 1)\sim
(1-{\cal T})^{-\frac{1}{3}}$. For $\zeta>1$,
we find from Eq.~(\ref{eq:16}) that $\phi$ corresponding to ${\cal T}_{\rm max}$ is
${\rm arccosh}\, \zeta-\sqrt{1-\zeta^{-2}}$. This gives
\begin{equation}\label{eq:35}
    {\cal T}_{\rm max}=\frac{2}{1+\zeta \cosh\sqrt{1-\zeta^{-2}} -\sqrt{\zeta^2-1}\sinh \sqrt{1-\zeta^{-2}}}.
\end{equation}
For $\zeta\gg 1$, ${\cal T}_{\rm max}=2e/\zeta$.
In SM,
we show further that if the input
surface is reflective, the transition may occur also when the internal reflection
at the output surface passes a certain critical value,
at which $z_b$ at the output surface equals the sum of $z_b$ at the input surface
and $L$.

\begin{figure}
 \begin{center}
 \includegraphics[width=8.0cm]{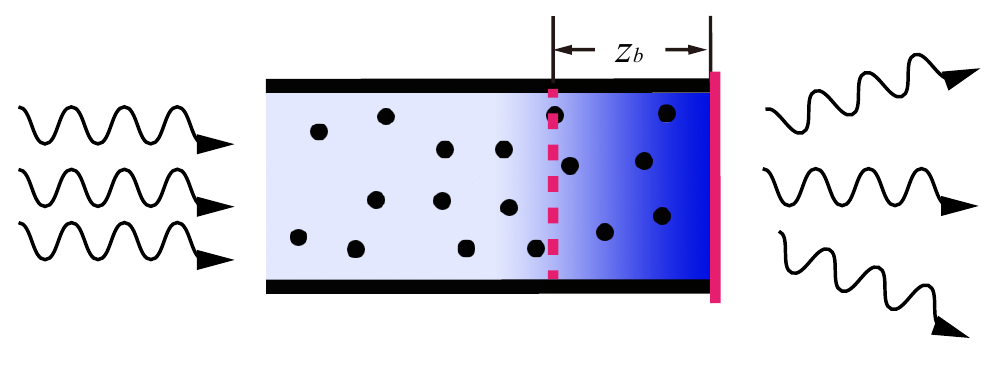}
\end{center}
 \caption{A resonant cavity formed by internal reflection at the
 output surface (red solid line) and the `virtual' reflector (red dashed line) formed by scatterers inside the medium.
 The latter is placed at a distance of $z_b$ from the output surface.}
 \label{fig:mirror}
\end{figure}

We consider a possible qualitative physical picture for
the transition. We suppose that internal reflection appears only on
the output surface. Waves that flow
from the input to the output surface
are reinjected into the sample as they reflect from the output surface. These
waves propagate diffusively over a typical distance $z_b$,
as implied by the boundary constraint for the $Q$ field,
which encodes information on the spatial variation of the intensity.
The plane which is located inside the medium
and at a distance $z_b$ from the output surface
may be viewed as a virtual reflector (see Fig.~\ref{fig:mirror}).
A `virtual' resonant cavity is thereby formed giving rise to perfect transmission.
(The same arguments apply if the input surface is reflective instead.)
For large internal reflection, $\zeta\geq 1$, the
virtual reflector falls outside the medium, and perfect transmission
would no longer be possible.

In summary, we have developed a supersymmetric field theory
to study the impact of surface interaction on the transmission eigenvalue statistics
in random media with arbitrary strength of disorder.
Although we focused here on quasi-one-dimensional systems and weak disorder, the
theory is also applicable to study the impact of surface reflection
in higher dimensions and strongly disordered environments.
The analytical calculations of the transition
in the distribution of high transmission eigenvalues
as the internal reflection at the output surface increases
are confirmed numerically.
The results are coherent effects for diffusive waves
that emerge from the wave nature
of propagation but are unrelated to weak localization.
They are therefore relevant to transmission in other wave systems
in a variety of applications such as,
for example, thermal phonon transport in dissimilar solids \cite{Little59}
where effects of surface reflection such as the Kapitza resistance
are not negligible.

We would like to thank Zhou Shi for advice on the simulation program and Leonid I. Glazman for useful discussions.
This work is supported by the NSF (No. DMR-1207446), by the NSFC (No. 11174174) and
by the Tsinghua University ISRP (No. 2011Z02151).

\newpage

\begin{widetext}

\begin{center}

{\bf Supplementary materials for:\\
Transmission eigenvalues in random media with surface reflection}\\

{\small Xiaojun Cheng,$^{1,2}$ Chushun Tian,$^{3}$ and Azriel Z. Genack$^{1,2}$\\
{\it $^1$Department of Physics, Queens College,
City University of New York, Flushing, New York 11367, USA\\
$^2$The Graduate Center, The City University of New York, New York, NY 10016 USA\\
$^3$Institute for Advanced Study, Tsinghua University,
Beijing 100084, China
}}

\end{center}

We present below derivations of some of the results in this paper.\\
\\
\noindent{\bf I. Field theoretic formalism.}\\

We introduce the generating function
\begin{equation}\label{eq:4}
    {\cal Z}(\theta,\phi)=\left\langle \frac{{\rm det}(1-\gamma_1\gamma_2 {\bf t}{\bf t}^\dagger)}
    {{\rm det}(1-\zeta_1\zeta_2 {\bf t}{\bf t}^\dagger)}\right\rangle,
\end{equation}
with the parameters $\gamma_1=\frac{1}{2}\sin\theta,\, \gamma_2=\tan\frac{\theta}{2},\,
\zeta_1=\frac{i}{2}\sinh\phi$ and $\zeta_2=i\tanh\frac{\phi}{2}$.
Taking the derivative of ${\cal Z}$ with respect to $\zeta_2$, we find
\begin{equation}\label{eq:13}
    F(\phi)=-\partial_{\zeta_2}{\cal Z}|_{\theta=i\phi}.
\end{equation}
The great advantage of introducing the generating function is that
it can be related to the
retarded (advanced) Green function, defined as
$(\omega_\pm^2- {\hat H} )G^{R,A}_{\omega^2}({\bf r},{\bf
 r}')=\delta({\bf r} -{\bf r}')$,
which describe wave propagation in random dielectric media on the microscopic level.
Here, the wave group velocity (in surrounding air) is set to unity, $\omega$ is
the circular wave frequency, and $\omega_\pm =
\omega \pm i\delta$ with $\delta$ positive infinitesimal. The `Hamiltonian' is
${\hat H}\equiv -\nabla^2 -\omega^2 \delta \epsilon({\bf r})$, where $\delta \epsilon$
represents the fluctuations in the dielectric function.
We consider two virtual parallel cross sections inside the random medium
and the energy current flowing between.
The moments of the transmission matrix ${\bf t}{\bf t}^\dagger$
can then be expressed in
terms of the retarded (advanced) Green function via
${\rm tr}({\bf t}{\bf t}^\dagger)^n =
    {\rm tr}(\hat j \delta_{L}G^{A}_{\omega^2}\hat j \delta_{R}G^{R}_{\omega^2})^n
    $,
where $\hat j$ is the energy flux operator,
and $\delta_{L (R)}$ restricts the spatial integral on the left (right) cross section.
Combined with the identity: ${\rm det}(1-x{\bf t}{\bf t}^\dagger) \equiv
\exp[-\sum_{n=1}^\infty \frac{x^n}{n}{\rm tr}({\bf t}{\bf t}^\dagger)^n]$,
this gives: ${\rm det}(1-x{\bf t}{\bf t}^\dagger)=
{\rm det} (1-x\hat j \delta_{L}G^{A}_{\omega^2}\hat j \delta_{R}G^{R}_{\omega^2})$.
As a result,
\begin{equation}\label{eq:33}
    {\cal Z}(\theta,\phi)=\left\langle \frac{{\rm det}(1-\gamma_1\gamma_2
    \hat j \delta_{L}G^{A}_{\omega^2}\hat j \delta_{R}G^{R}_{\omega^2})}
    {{\rm det}(1-\zeta_1\zeta_2
    \hat j \delta_{L}G^{A}_{\omega^2}\hat j \delta_{R}G^{R}_{\omega^2})}\right\rangle.
\end{equation}

The structure of the ratio of two functional determinants invites the
application of the supersymmetric technique of Efetov \cite{Efetov97}.
Procedures for applying this technique to classical waves
in open media are reviewed in Ref.~\cite{Tian13}. Here, we only
outline the scheme. First, we introduce a $4$-component supervector
$\psi =\{\psi_{\lambda\alpha}\}$, where
the index $\lambda=1,2$ distinguishes distinct analytic structure of the advanced (retarded) Green function,
and
$\alpha={\rm f},{\rm b}$ the fermionic (bosonic) variables.
Recall that in this work we ignore time-reversal symmetry.
Next, we perform the disorder averaging,
which introduces the effective interaction of the supervector field.
We then adopt the super-Hubbard-Stratonovich transformation \cite{Efetov97} to decouple
the interaction, which introduces a $4\times 4$ supermatrix field,
$Q$. After separating the slow and fast modes, we
obtain a nonlinear supermatrix $\sigma$ model. As a result, a
general physical observable is expressed in terms of the functional
integral over the $Q$-field which obeys the nonlinear constraint: $Q^2=1$.
In the last step, we derive the boundary condition satisfied by the $Q$-field.
It is very important that this accounts for the open nature of random media,
and retains the distinction in localization physics between open media
and infinite (closed) media. As a result,
\begin{equation}\label{eq:34}
    {\cal Z}(\theta,\phi)=\int D[Q] e^{-{\cal F}[Q]},\quad
{\cal F}[Q]=
\frac{\xi}{8}
\int_0^L dx {\rm str} (\partial_x Q-i[{\cal A}_x,Q])^2.
\end{equation}
Here,
the `gauge field'
${\cal A}_x=\delta_R (\gamma_1 \Bbb{E}^{12}_{{\rm ff}}\oplus\zeta_1 \Bbb{E}^{12}_{\rm bb})
\oplus \delta_L (\gamma_2 \Bbb{E}^{21}_{\rm ff}\oplus\zeta_2 \Bbb{E}^{21}_{\rm bb})$
which vanishes at the interfaces.
$\Bbb{E}^{\lambda\lambda'}_{\alpha\alpha'}$ is a projector which takes the value of unity
for the entry $(\lambda\alpha,\lambda'\alpha')$ and is zero otherwise.
The supermatrix field is constrained by
the boundary condition arising from internal reflection,
\begin{equation}
(2z_b
Q
\partial_x Q
\pm [Q,\Lambda])|_{x=0,L}=0,
\label{eq:6}
\end{equation}
where $z_b=0.7\ell\frac{1+R}{1-R}$ depends on
the internal reflection coefficient $R$ at the interface,
and the $+$ ($-$) sign
corresponds to the left (right) interface located at
$x=0$ ($x=L$). Notice that Eq.~(\ref{eq:6}) is the source of the difference between the
present and previous theories \cite{Rejaei96}.

Substituting Eq.~(\ref{eq:34}) into Eq.~(\ref{eq:13}), we find
$F(\phi)=-\frac{i}{2}\xi
\int D[Q]{\rm str}(\delta_L \Bbb{E}^{21}_{\rm bb}Q(\partial_x Q-
i[{\cal A}_x,Q])) e^{-{\cal F}[Q]}|_{\theta=i\phi}$. Notice that
the spatial integral included by the supertrace is effectively restricted on the
left cross section by $\delta_L$.
Making the gauge transformation:
$Q(x) \rightarrow S(x) Q(x) S^{-1}(x),\,S(x)={\rm T}e^{i\int_{-\infty}^x {\cal A}_x dx'}$, with
`T' denoting the path-ordered product, we obtain
Eq.~(4).\\

\noindent {\bf II. Derivations of Eqs.~(7).
}\\

The solution of $\Phi(x),\Theta(x)$ must satisfy two
conditions: it must be well defined for arbitrary
$\zeta\geq 0$ and it must converge to its limiting case at $\zeta=0$ given e.g., before in
Ref.~\cite{Zirnbauer04}. From the boundary constraints on $\Phi$ given in
Eqs.~(6a) and (6b),
we obtain
\begin{equation}\label{eq:18}
    -\sinh \phi_0=\sinh(C_\phi+\phi_0-\phi),
\end{equation}
giving $\phi_0=-(C_\phi-\phi)/2$. Inserting this back into Eq.~(6a)
we obtain the first equation of (7).

From the boundary constraints of $\Theta$ given in
Eqs.~(6a) and (6b),
we obtain
\begin{equation}\label{eq:19}
    -\sin \theta_0=\sin(C_\theta+\theta_0-\theta).
\end{equation}
There are two possibilities for matching both sides. (i) $(2n+1)\pi + \theta_0 =
C_\theta+\theta_0-\theta,n\in \mathbbm{N}$ which gives
$C_\theta=\theta+(2n+1)\pi$.
However, this solution must be discarded because it cannot match
the limiting case of $\zeta=0$, where $C_\theta=\theta+2n\pi$ \cite{Zirnbauer04}.
(ii) $-\theta_0=C_\theta+\theta_0-\theta-2n \pi$
which gives
\begin{equation}\label{eq:20}
    \theta_0 =-\frac{C_\theta-\theta}{2}+n\pi.
\end{equation}
Inserting this back into Eq.~(6a)
we obtain
\begin{equation}\label{eq:21}
    \zeta C_\theta +\sin \left(\frac{C_\theta-\theta}{2}-n\pi\right)=0.
\end{equation}
Suppose that $n$ is odd. It is easy to see that Eq.~(\ref{eq:21}) has
a unique real solution for arbitrary $\zeta\geq 0$. In addition,
this solution satisfies $0<\frac{\theta-C_\theta}{2}<\pi$. Because of $0<\theta_0<\pi$
Eq.~(\ref{eq:20}) requires $n=0$. This contradicts the presumed odd parity of
$n$ and therefore $n$ must be even. We thereby obtain the second equation
of (7).
In fact, it is easy to see that the latter has unique real solution
for arbitrary $\zeta\geq 0$. In addition, this solution satisfies $0<\frac{\theta-C_\theta}{2}<\pi$
which implies $n=0$. Thus, we complete the proof of Eq.~(7).
\\

\noindent {\bf III. Derivations of Eq.~(2).
}\\

From the first equation of (7)
we obtain
\begin{equation}\label{eq:22}
    \zeta C_{\phi_\pm}=-\sinh \frac{C_{\phi_\pm}-\phi_\pm}{2}.
\end{equation}
For the moment we assume $\phi_\pm>0$ and set these to $\phi\pm i\pi$ in the final results.
From this equation we further obtain
\begin{eqnarray}
\label{eq:23}
  \zeta (C_{\phi_+}+C_{\phi_-}) &=& -2 \sinh \left[\frac{1}{2}
  \left(\frac{C_{\phi_+}+C_{\phi_-}}{2}-{\bar \phi}\right)\right]
  \cosh \left(\frac{1}{2}\frac{C_{\phi_+}-C_{\phi_-}-\Delta \phi}{2}\right),\nonumber\\
  \zeta (C_{\phi_+}-C_{\phi_-}) &=& -2 \cosh \left[\frac{1}{2}
  \left(\frac{C_{\phi_+}+C_{\phi_-}}{2}-{\bar \phi}\right)\right]
  \sinh \left(\frac{1}{2}\frac{C_{\phi_+}-C_{\phi_-}-\Delta \phi}{2}\right).
\end{eqnarray}
where ${\bar \phi} \equiv (\phi_++\phi_-)/2,\,
\Delta \phi \equiv \phi_+-\phi_-$. These two equations may be rewritten as
\begin{eqnarray}
\label{eq:31}
  \frac{C_{\phi_+}+C_{\phi_-}}{2}-{\bar \phi} &=& -\frac{1}{\zeta}\sinh \left[\frac{1}{2}\left(\frac{C_{\phi_+}+C_{\phi_-}}{2}-{\bar \phi}\right)\right]
  \cosh \left(\frac{1}{2}\frac{C_{\phi_+}-C_{\phi_-}-\Delta\phi}{2}\right) - {\bar \phi}, \nonumber\\
  \frac{C_{\phi_+}+C_{\phi_-}}{2}-{\bar \phi} &=&
  \pm 2{\rm arccosh} \left[-\frac{\zeta (C_{\phi_+}-C_{\phi_-})/2}{\sinh \left(\frac{1}{2}\frac{C_{\phi_+}-C_{\phi_-}-\Delta\phi}{2}\right)}\right].
\end{eqnarray}
The sign on the right-hand side of the second equation will be determined below.
Inserting the second equation into the first gives
\begin{eqnarray}
\label{eq:32}
  \pm 2{\rm arccosh} \left[-\frac{\zeta (C_{\phi_+}-C_{\phi_-})/2}{\sinh \left(\frac{1}{2}\frac{C_{\phi_+}-C_{\phi_-}-\Delta\phi}{2}\right)}\right]
  =\mp \frac{1}{\zeta}\sqrt{\left[\frac{\zeta (C_{\phi_+}-C_{\phi_-})/2}{\sinh \left(\frac{1}{2}\frac{C_{\phi_+}-C_{\phi_-}-\Delta\phi}{2}\right)}\right]^2-1}
  \cosh \left(\frac{1}{2}\frac{C_{\phi_+}-C_{\phi_-}-\Delta\phi}{2}\right)-{\bar \phi}.
\end{eqnarray}
Upon setting ${\bar \phi}=\phi,\,\Delta \phi=2i\pi$, we have
$f(\phi)=(C_{\phi_+}-C_{\phi_-})/(2i\pi)$. Inserting these into Eq.~(\ref{eq:32}), we find
\begin{eqnarray}
\label{eq:24}
  \mp\phi = 2{\rm arccosh}\frac{\pi\zeta f}{\cos\frac{\pi f}{2}}
  +\frac{\sin\frac{\pi f}{2}}{\zeta}
  \sqrt{\left(\frac{\pi\zeta f}{\cos\frac{\pi f}{2}}\right)^2-1}.
\end{eqnarray}
Because $\phi>0$ we need to take the positive sign and therefore obtain Eq.~(2).
\\
\\
\noindent {\bf IV. Derivations of Eq.~(3).
}\\

First of all, thanks to $\ell/L\ll 1$ we may simplify Eqs.~(8a) and (8b)
to
\begin{subequations}
\begin{equation}\label{eq:44}
    \zeta C_\phi=-\sinh \left(C_\phi-\phi\right),
\end{equation}
\begin{equation}\label{eq:45}
    \zeta C_\theta=-\sin \left(C_\theta-\theta\right).
\end{equation}
\end{subequations}
Then, the procedure is similar to the derivations of Eq.~(2).
From Eq.~(\ref{eq:44}) we obtain
\begin{equation}\label{eq:27}
    \zeta C_\pm=-\sinh (C_\pm-\phi_\pm).
\end{equation}
Here again, we assume for the moment $\phi_\pm>0$ and set these to $\phi\pm i\pi$ in the final results.
From Eq.~(\ref{eq:27}), we obtain
\begin{eqnarray}
\label{eq:28}
  \zeta (C_{\phi_+}+C_{\phi_-}) &=& -2 \sinh \left(\frac{C_{\phi_+}+C_{\phi_-}}{2}-{\bar \phi}\right)
  \cosh \frac{C_{\phi_+}-C_{\phi_-}-\Delta\phi}{2}, \nonumber\\
  \zeta (C_{\phi_+}-C_{\phi_-}) &=& -2 \cosh \left(\frac{C_{\phi_+}+C_{\phi_-}}{2}-{\bar \phi}\right)
  \sinh \frac{C_{\phi_+}-C_{\phi_-}-\Delta\phi}{2}.
\end{eqnarray}
These two equations may be rewritten as
\begin{eqnarray}
\label{eq:29}
  \frac{C_{\phi_+}+C_{\phi_-}}{2}-{\bar \phi} &=& -\frac{1}{\zeta}\sinh \left(\frac{C_{\phi_+}+C_{\phi_-}}{2}-{\bar \phi}\right)
  \cosh \frac{C_{\phi_+}-C_{\phi_-}-\Delta\phi}{2} - {\bar \phi}, \nonumber\\
  \frac{C_{\phi_+}+C_{\phi_-}}{2}-{\bar \phi} &=&
  \pm {\rm arccosh} \left(-\frac{\zeta (C_{\phi_+}-C_{\phi_-})/2}{\sinh \frac{C_{\phi_+}-C_{\phi_-}-\Delta\phi}{2}}\right).
\end{eqnarray}
The sign on the right-hand side of the second equation will be determined below.
Inserting the second equation into the first gives
\begin{eqnarray}
\label{eq:30}
  \pm {\rm arccosh} \left(-\frac{\zeta (C_{\phi_+}-C_{\phi_-})/2}{\sinh \frac{C_{\phi_+}-C_{\phi_-}-\Delta\phi}{2}}\right)
  =\mp \frac{1}{\zeta}\sqrt{\left(\frac{\zeta (C_{\phi_+}-C_{\phi_-})/2}{\sinh \frac{C_{\phi_+}-C_{\phi_-}-\Delta\phi}{2}}\right)^2-1}
  \cosh \frac{C_{\phi_+}-C_{\phi_-}-\Delta\phi}{2}-{\bar \phi}.
\end{eqnarray}
Setting ${\bar \phi}=\phi,\,\Delta \phi=2i\pi$, we obtain
\begin{eqnarray}
\label{eq:30}
  \mp \phi = {\rm arccosh} \frac{\pi\zeta f}{\sin \pi f}
  -\frac{\cos\pi f}{\zeta}\sqrt{\left(\frac{\pi\zeta f}{\sin \pi f}\right)^2-1}.
\end{eqnarray}
Because $\phi>0$ we need to take the positive sign and therefore obtain Eq.~(3).
\\
\\
\noindent {\bf V. Transition for imbalanced internal reflection.}\\

Here we show that if the internal reflection at the input
surface, $R_1>0$, is fixed and the internal reflection at the output
surface, $R_2$, increases, the distribution of transmission eigenvalues
may also exhibit a transition. For convenience below, we introduce the
following parameters: $z_{b1,b2}=0.7\ell (1+R_{1,2})/(1-R_{1,2})$ and
$\zeta_{1,2}= z_{b1,b2}/L$. To simplify technical discussions below,
we assume $\zeta_1$ small and arbitrary $\zeta_2$.

\begin{figure}
 \begin{center}
 \includegraphics[width=10.0cm]{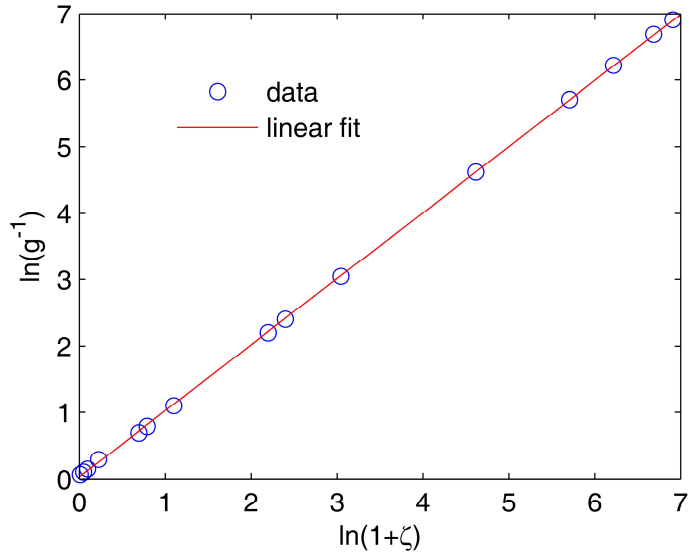}
\end{center}
 \caption{The average conductance obeys Ohm's law. The data can be 
 fitted to a straight line whose slope is unity.}
 \label{fig:Ohm}
\end{figure}

For imbalanced internal reflection, $z_{b1}\neq z_{b2}$,
the boundary conditions (6a) and (6b)
are replaced by
\begin{eqnarray}
(z_{b1}\partial_x \Phi -\sinh \Phi)|_{x=0} =0,\qquad
(z_{b2}\partial_x \Phi +\sinh (\Phi-\phi))|_{x=L}=0.
\label{eq:37}
\end{eqnarray}
Here, we present only the boundary conditions of $\Phi(x)$ because, as
we have seen above, they are sufficient for the purpose of calculating the deviation factor $f({\cal T})$.
Inserting the solution to Eq.~(6a),
i.e., $\Phi(x)=C_\phi x/L + \phi_0$, into Eq.~(\ref{eq:37}), we obtain
\begin{eqnarray}
\zeta_{1}C_\phi - \sinh \phi_0 =0,\qquad
\zeta_{2}C_\phi +\sinh (C_\phi + \phi_0-\phi)=0.
\label{eq:38}
\end{eqnarray}
Because of $\zeta_1\ll 1$ we may expand the sinh function in the first equation.
We then eliminate $\phi_0$ to obtain
\begin{eqnarray}
\zeta_{2}C_\phi =-\sinh ((1+\zeta_1) C_\phi-\phi).
\label{eq:39}
\end{eqnarray}
Defining $\tilde C_\phi\equiv (1+\zeta_1) C_\phi$ and $\tilde \zeta\equiv \zeta_2/(1+\zeta_1)$,
we rewrite this equation as
\begin{eqnarray}
\tilde \zeta\tilde C_\phi =-\sinh (\tilde C_\phi-\phi).
\label{eq:40}
\end{eqnarray}
The deviation factor is $f=
(\tilde C_{\phi+i\pi}-\tilde  C_{\phi-i\pi})/[2i\pi(1+\zeta_1)]
\equiv \tilde f/(1+\zeta_1)$, and the distribution of transmission eigenvalues is
\begin{eqnarray}
\rho({\cal T})=\tilde f({\cal T})\times\left[
    \frac{\xi}{2(L+z_{b1})} \frac{1}{{\cal T}\sqrt{1-{\cal T}}}\right].
\label{eq:41}
\end{eqnarray}
Now we may repeat the procedure of Sec. IV. As a result, we obtain the same equation
as (\ref{eq:30}) albeit with the replacement: $\zeta\rightarrow \tilde \zeta,\,
f \rightarrow \tilde f$. This implies that the same transition as that discussed in the main
text also occurs here. The transition point is
$\tilde \zeta=1$, i.e., $z_{b2}=L+z_{b1}$.
\\
\\
\noindent {\bf VI. Average conductance.}\\

Here we only study the most interesting case in which a single surface is reflective.
We first rescale the average conductance by $\xi/L$. The resultant quantity, denoted as $g$, is given by
\begin{equation}\label{eq:50}
    g=\frac{1}{2}\int_0^1 d{\cal T} {\cal T} \times \frac{f({\cal T})}{{\cal T}\sqrt{1-{\cal T}}}
    =\frac{1}{2}\int_0^1 d{\cal T} \frac{f({\cal T})}{\sqrt{1-{\cal T}}},
\end{equation}
which depends on a single parameter $\zeta$.
We cannot analytically calculate $g(\zeta)$.
Rather, we numerically compute it for a wide range of $\zeta$
from $10^{-1}$ to $10^3$. We find that the numerical results for
$g(\zeta)$ are well fitted by $1/(1+\zeta)$, as shown in Fig.~\ref{fig:Ohm}.
Therefore, the average conductance is $g(\zeta)\xi/L =\xi/(L+z_b)$.
It is important that in spite of the transition
exhibited by the distribution $\rho({\cal T})$ at the critical value of
$\zeta=1$, the average conductance obeys Ohm's law for both weak ($\zeta\ll 1$)
and strong ($\zeta\gg 1$) internal reflection.

\end{widetext}


\begin{thebibliography}{}
\bibitem{note_extrapolation} This entails the standard terminology namely the
extrapolation length for $z_b$. However, this is very misleading
in many contexts. For example, for localized waves,
the definition of $z_b$ in the diffusion model
or radiative transfer theory is meaningless because these theories cease to work.
Instead, it has been found \cite{Tian08,Tian13} that in the field theory of localization in open media, exactly the same relation,
$z_b=0.7\ell (1+R)/(1-R)$, enters into the boundary condition satisfied by the field.
For diffusive waves, the field-theoretic boundary condition is equivalent to that
in the diffusion model or radiative transfer theory.
\end{thebibliography}
\end{document}